\documentclass[11pt]{llncs}
\usepackage{fullpage}

\newcommand{\Oh}[1]
    {\ensuremath{\mathcal{O}\!\left( {#1} \right)}}
\newcommand{\occ}
    {\ensuremath{\mathsf{occ}}}

\begin{document}

\title{A Compressed Self-Index\\
    for Genomic Databases}
\author{Travis Gagie\inst{1}
    \and Juha K\"arkk\"ainen\inst{2}
    \and\\ Yakov Nekrich\inst{3}
    \and Simon J. Puglisi\inst{4}}
\institute{Aalto University, Finland
    \and University of Helsinki
    \and University of Bonn
    \and King's College, London}
\maketitle

\begin{abstract}
Advances in DNA sequencing technology will soon result in databases of thousands of genomes.  Within a species, individuals' genomes are almost exact copies of each other; e.g., any two human genomes are 99.9\% the same.  Relative Lempel-Ziv (RLZ) compression takes advantage of this property: it stores the first genome uncompressed or as an FM-index, then compresses the other genomes with a variant of LZ77 that copies phrases only from the first genome.  RLZ achieves good compression and supports fast random access; in this paper we show how to support fast search as well, thus obtaining an efficient compressed self-index.
\end{abstract}

\section{Introduction} \label{sec:introduction}

DNA sequencing technology has advanced to the point that in the foreseeable future it will be practical for many people to pay for copies their genomes~\cite{TGP,PGP}.  This raises the question how to store many individual genomes compactly but such that we can still search them quickly.  Any two human genomes are 99.9\% the same, but compressed self-indexes based on compressed suffix arrays, the Burrows-Wheeler Transform or LZ78 (see~\cite{NM07} for a survey) do not take full advantage of this similarity.  Kreft and Navarro~\cite{KN11} recently introduced a compressed self-index based on LZ77, which compresses repetitive sequences very well; however, two drawbacks of their index is that i) it uses a lot of space for the first genome and ii) the search time depends on the depth of nesting of the LZ77 parse.  A compressed self-index stores a string \(S [1..n]\) in compressed form such that, given a pattern \(P [1..m]\), we can quickly list the $\occ$ occurrences of $P$ in $S$.  In this paper we show how to build a compressed self-index on top of Relative Lempel-Ziv (RLZ) compression, which allows us to i) store the first, reference genome separately and more compactly, ii) restrict the depth of nesting to 1 and iii) reduce the dependence on the pattern length from quadratic to linear in the search time.

Kuruppu, Puglisi and Zobel~\cite{KPZ10} introduced RLZ specifically for compression of genomic databases.  Their idea is to store the first genome either uncompressed or as an FM-index, then compresses the other genomes with a variant of LZ77 that copies phrases only from the first genome --- thus restricting the depth of nesting in the parse to 1.  Although they showed that RLZ compresses genomic databases well and supports fast random access, they did not show how to support search in the whole database.  We show how, if $G$ is the reference genome of length $n$, $T$ is the rest of the database and the RLZ parse of $T$ with respect to $G$ has $r$ phrases, then we can store the whole database in \((1 + 1 / \epsilon) n H_k (G) + \Oh{r (\log n + \log^{1+\epsilon} r)} + \Oh{n}\) bits such that, given a pattern $P$ of length $m$, in \(\Oh{(m + \occ_0) \log^\epsilon n + \occ_1 + \occ_2}\) time we can list the $\occ_0$ occurrences of $P$ in $G$ and the \(\occ_1 + \occ_2\) occurrences of $P$ in $T$.

\section{The Data Structure} \label{sec:results}

Let $G[1..n]$ be the reference sequence and let $T[1..N]$ be the rest
of the text.  We assume that the alphabet $\Sigma$ has
a polylogarithmic size in $n$.  The RLZ-parse of $T$ with respect to
$G$ is the partitioning $T_1T_2..T_r$ of $T$ into $r$ phrases such
that $T_i$ is the longest prefix of $T_i..T_r$ that occurs in $G$.
Let $D$ be the set of $d$ distinct phrases in the parse.

We divide the problem of finding the occurrences of a pattern \(P[1..m]\)
into three phases: (i) finding the $\occ_0$ occurrences in $G$, (ii)
finding the $\occ_2$ secondary occurrences in $T$, i.e., occurrences
that are completely contained in a single phrase, and (iii) finding
the $\occ_1$ primary occurrences in $T$, i.e., occurrences that cross
a phrase boundary.

We store the reference sequence $G$ in an compressed suffix array (CSA) of \((1 / \epsilon) n H_k + \Oh{n}\) bits~\cite{GGV03}, where $\epsilon$ is an arbitrary constant with \(0 < \epsilon \leq 1\).
Then we can find all the $occ_0$ occurrences of a pattern $P[1..m]$ in \(\Oh{m \log \sigma + \occ_0 \log^\epsilon n}\) time.

To support search for secondary occurrences, we build a data structure
for 2-dimensional, 2-sided range-reporting on an \(n \times n\) grid,
on which we place $r$ points, with each point \((i, j)\) indicating
that a phrase is copied from \(G [i..j]\).  Since the queries will be
2-sided --- in particular, given a query point \((x, y)\) we want to
find all the points \((i, j)\) on the grid such that \(i \leq x\) and
\(j \geq y\), and the associated phrase, whose source \(G [i..j]\)
includes \(G [x..y]\) --- this data structure can be implemented with
a predecessor data structure for the points' horizontal coordinates, a
range-maximum for their vertical coordinates (sorted by their
horizontal coordinates) and a map from points to text positions.
These data structures take \(\Oh{r \log (n + r)} + o (n)\) bits and answer
queries in $\Oh{p}$ time, where $p$ is the number of points returned.
For each occurrence \(G [x..y]\) of $P$ in $G$, we query the 2-sided
range-reporting data structure to find all $p$ phrases whose sources
include \(G [x..y]\), in $\Oh{p}$ time.

\begin{lemma} \label{lem:contained}
There is a data structure of \(\Oh{r \log (n + r)} + o (n)\) bits such
that, given the $\occ_0$ occurrences of $P$ in $G$, in $\Oh{\occ_0+\occ_2}$
time we can list the $\occ_2$ secondary occurrences of $P$ in $T$.
\end{lemma}

To find primary occurrences, we represent each phrase boundary with a
pair \((i, j)\), where $i$ is the rank of the suffix of $T$ starting
at the phrase boundary in the lexicographical ordering of all suffixes
of $T$ starting at phrase boundaries, and $j$ is the rank of the
phrase ending at the phrase boundary in the lexicographical ordering
of $D^R$, the set of reversed distinct phrases.  We store the pairs in
a data structure for 2-dimensional range reporting by Alstrup et
al.~\cite{ABR00}, which requires $\Oh{r \log^{1+\epsilon} r}$ bits of
space and answers queries in $\Oh{\log \log r + p}$ time, where $p$ is
the number of points returned. Given the lexicographical interval of
suffixes of $T$ starting at phrase boundaries and having $P[i..m]$ as
a prefix and the lexicographical interval of $D^R$ of reversed phrases
having \((P[1..i-1])^R\) as a prefix, we can find all primary
occurrences of $P$ such that first phrase boundary inside the
occurrence is at position $i$. The rest of the section describes how
these intervals can be computed efficiently.

We augment the CSA of $G$ with $\Oh{n}$-bit data structures storing the longest common prefix
(LCP) array~\cite{Fis10} and its next/previous-smaller-value (NSV/PSV) array~\cite{Fis11}.
Then, for any string $X$ and symbol $c$,
given the lexicogaphical interval of suffixes of $G$ beginning with
$X$ we can compute the interval for $cX$ in $\Oh{\log \sigma}$ time, and given
the interval for $Xc$ we can compute the interval for $X$ in $\Oh{\log^\epsilon n}$ time.
Given a pattern $P[1..m]$, for all \(i \in [1..m]\), let
$\ell(i)$ be an integer such that \(P[i..\ell(i)-1]\) is the longest
prefix of $P[i..m]$ that occurs in $G$. We can compute $\ell(i)$ for
all $i$ in $\Oh{m \log^\epsilon n}$ time using the CSA and the LCP and NSV/PSV
data structures.  Then, for some $k$, \(P[i..\ell(i)-1]
P[\ell(i)..\ell^2(i)-1] \ldots P[\ell^k(i)..m]\) is the RLZ-parse of
$P[i..m]$ with respect to $G$. If $P[i..m]$ occurs in $T$ starting at
a phrase boundary, then the first $k$ phrases following that phrase
boundary must match \(P[i..\ell[i]-1], P[\ell[i]..\ell^2[i]-1],
\ldots, P[\ell^{k-1}[i]..\ell^k[i]-1]\) exactly, and the $(k+1)$st
phrase must begin with \(P[\ell^k[i]..m]\).

When computing $\ell(i)$, we also obtain the lexicographical interval
of the suffixes of $G$ that begin with \(P[i..\ell(i)-1]\), and the
next step is to turn these into lexicographical ranks and intervals in
$D$, the set of distinct phrases. We represent each phrase in $D$ with
the integer $qn+k$, where $q$ is the lexicographical rank of the
smallest of $G$ that begins with the phrase, and $k$ is the length of
the phrase. Note that the integer is a complete description of the
phrase, given $G$, and it is consistent with the lexicographical
ordering of the phrases.  We store the integers in a data structure of
\(\Oh{d \log n}\) bits supporting \(\Oh{\log\log d}\) time predecessor
queries.  If $[p..q]$ is the lexicograpical interval of suffixes of
$G$ that begin with \(P[i..\ell(i)-1]\), then \([pn+\ell(i)-i..qn+n]\)
is the interval of phrases beginning with \(P[i..\ell(i)-1]\).  If the
first phrase in that interval is \(pn+\ell(i)-i\), then it matches
\(P[i..\ell(i)-1]\) exactly.

Let $R$ be the representation of the text $T$ as a sequence of
phrases, with each phrase represented by its rank in the
lexicographical ordering of $D$. The suffixes of $R$ correspond to
suffixes of $T$ starting at phrase boundaries.  We store a predecessor
data structure of \(\Oh{r+d+o(r)}\) bits, which maps a phrase in $D$
into the lexicographical interval of suffixes of $R$ that begin with
that phrase.  We also store the FM-index~\cite{FM05,FMMN07} of $R$ in $\Oh{r\log d}$ bits
supporting a backward search step in \(\Oh{\log\log d}\) time.  Let
$h$ be the smallest integer such that $P[h..m]$ occurs in $G$.  For
all $i \in [h..m]$, we map the interval of phrases beginning with
\(P[i..m]\) into the interval of suffixes of $R$ beginning with those
phrases.  Then, for \(i=h-1,\ldots,1\), having the interval of $R$'s
suffixes for \(P[\ell(i)..m]\) and the rank of the phrase matching
\(P[i..\ell(i)-1]\), we can compute the interval for \(P[i..m]\) with
one backward search step. Thus the intervals for \(P[i..m]\) for all
$i$ can be computed in \(m\log\log d\) time.

Finally, to find reversed phrases beginning with \((P[1..i-1])^R\), we
store an FM-index for the reverse $G^R$ of $G$ in \(n H_k (G) + o (n)\) bits.
We use the FM-index to
compute the lexicographical interval of suffixes of $G^R$ that begin
with \((P[1..i-1])^R\) for all $i\in[1..m]$ in $\Oh{m}$ time.  Using
the technique of representing phrases with integers as above, we can
store $D^R$ in \(\Oh{d\log n}\) bits so that the interval of reversed
phrases beginning with \((P[1..i-1])^R\) can be computed in
\(\Oh{\log\log d}\) time.

\begin{theorem} \label{thm:main}
We can store a reference sequence $G$ of length $n$ and a text $T$,
whose RLZ-parse with respect to $G$ has $r$ phrases, in
\[(1 + 1 / \epsilon) n H_k (G) + \Oh{r (\log n + \log^{1+\epsilon} r)} + \Oh{n}\]
bits such that, given a pattern $P$ of length $m$, in
\[\Oh{(m + \occ_0) \log^\epsilon n + \occ_1 + \occ_2}\]
time we can list the $\occ_0$ occurrences of $P$ in $G$ and the \(\occ_1 + \occ_2\) occurrences of $P$ in $T$.
\end{theorem}

\bibliographystyle{plain}
\bibliography{rlz_index}

\end{document}